# Forged Channel: A Breakthrough Approach for Accurate Parkinson's Disease Classification using Leave-One-Subject-Out Cross-Validation

Hamidi, A.
Electerical Engineering Departement
K.N.Toosi Univeristy
Tehran, Iran
amirreza.hamidi@email.kntu.ac.ir

Mohamed-Pour, K.
Electerical Engineering Departement
K.N.Toosi Univeristy
Tehran, Iran
kmpour@eetd.kntu.ac.ir

Yousefi, M.
Electerical Engineering Departement
K.N.Toosi Univeristy
Tehran, Iran
mohammad.yousefi@email.kntu.ac.ir

*Abstract*— This paper introduces a novel technique called "Forged Channel," which aims to comprehensively represent EEG signals in order to achieve accurate classification of Parkinson's disease. The forged channel method prepares EEG signals in a manner that allows a deep learning model to effectively perceive all EEG channels within a single input. By employing this approach alongside a convolutional neural network, an impressive accuracy of 90.32% was achieved using leave-one-subject-out cross-validation. This performance closely reflects real-world conditions, highlighting the superiority of our method compared to similar approaches.

*Keywords*— *Parkinson's Disease (PD), Deep Learning, Convolutional Neural Network, SPWVD [1] Transformation, Electroencephalogram (EEG), Classification.*

## I. Introduction

Between 7 to 10 million people worldwide suffer from Parkinson's disease, which is a prevalent chronic neurological disorder that worsens over time [1]. The reason for its occurrence is the loss of dopamine-producing neurons in the substantia nigra region of the brain [2]. Symptoms of this disease include rigidity and tremors at rest [3], only appearing in the advanced stages of nerve damage, when the patient has lost 50 to 70 percent of their substantia nigra neurons [4] [5]. Diagnosis of this disease is usually based on clinical methods and by observing motor symptoms such as repetitive limb movement, resistance to involuntary movements, self-movement, balance, and walking patterns. In addition, diagnosing this disease requires an experienced physician, and there is still a possibility of error [6]. Therefore, advances in new diagnostic techniques can help detect this disease in its early stages and provide the possibility of prescribing dopaminergic drugs to improve the long-term quality of life of patients.

EEG signals are utilized to solve the problem of distinguishing between Healthy Control (HC) and Parkinson's Disease (PD) subjects. These signals contain valuable information that serves as a discriminative criterion. Various methods have been proposed to enhance the representation of EEG signals and improve classification accuracy using deep learning models. The objective of all these studies is to accurately classify HC and PD subjects by capturing features within EEG signals.

For example, Loh et al. [7] achieved an impressive accuracy of 99.46% by applying Gabor transformation using a convolutional neural network. Similarly, Khare et al. [8] obtained a remarkable 100% accuracy by using the SPWVD transformation instead of Gabor transformation. Lee et al. [9] employed segmented EEG signals to create a 3D input for a deep learning model, adjusting the dimensions based on segment length, number of segments, and channels. Other studies, such as Oh et al. [10] and Shi et al. [11], directly used segmented EEG signals as inputs for deep learning models. In these studies, the input was treated as a 2D matrix, with rows representing samples from individual channels and columns representing recordings from all channels at a specific time sample.

Now despite achieving a high accuracy score, these studies suffer from a limitation in their evaluation method: 10-fold and 5-fold cross-validation. Unfortunately, these approaches are unable to replicate real-world scenarios, unlike the more comprehensive leave-one-subject-out cross-validation. In our opinion, this problem arises due to the inability of their chosen method to represent the features of all EEG channels in a single input for the deep learning model. In other words, the deep learning model fails to look at the entirety of the all EEG channels features within a single input during test and train, resulting in a drop in accuracy during leave-one-subject-out cross-validation and a lack of generalizability to real-world situations. Bearing these concerns in mind, our study aims to address two primary objectives.

- Firstly, we aim to propose a method that can consolidate the features of all EEG channels into a single input. This approach will enable deep learning models to effectively utilize these optimized data and achieve improved performance overall.

- Secondly, we strive to attain exceptional accuracy through leave-one-subject-out cross-validation. By achieving high accuracy in this rigorous validation setting, we can ensure that our proposed model performs well in real-world scenarios and generalizes effectively beyond the testing phase.

The rest of the paper is organized as follows: Section 2, "Materials & Methodology," explains the dataset used, along with the preprocessing and preparation stages employed to clean and represent the EEG signals. Section 3 presents the results of leave-one-subject-out cross-validation, providing a detailed analysis for each subject in the dataset. In Section 4, the results are discussed and interpreted. Finally, the conclusion outlines the contributions of this study and its potential applications in real-world settings.

---

[1] Smoothed pseudo Wigner-Ville distribution.

## II. MATERIALS AND METHODOLOGY

This section is dedicated to detailing our work and the materials used in four distinct parts. In the first part, the specifics of our dataset will be elaborated upon. The second part will focus on the Pre-Processing stage and the various techniques utilized to clean our raw data. Subsequently, the third part will explain how our data was prepared and the innovative methods implemented to give us a richer representation of an EEG signal. Finally, the last section will outline the specifics of our deep learning model for classifying our PD and HC subjects.

### A. Dataset

The dataset utilized in this work was downloaded from Open Neuro [12]. This dataset comprises 16 HC and 15 PD subjects. The EEG signal of the PD subjects were recorded in two distinct scenarios; one scenario involved the use of medication, while the other scenario did not. For this study, we only considered the second scenario in which the subjects did not use medication. This dataset is a very popular and valid when studying Parkinson as it is very well distributed around different features. Also, in technical point of view, this dataset possesses two crucial features that hold significance in the subsequent stages. Firstly, the dataset maintains a sampling frequency of 512 Hz, ensuring high-resolution data acquisition. Secondly, each EEG recording within the dataset comprises of 32 channels, allowing for comprehensive analysis and exploration of neural activity.

### B. Pre-Processing

The Pre-Processing stage plays a crucial role in any EEG-related work due to the vulnerability of EEG signals to various types of noise. To ensure precise operations, a professional tool called EEG-LAB[13], available in MATLAB, is employed to help us with the pre-processing stage.

The Pre-Processing stage begins by passing the raw signal through a band-pass filter. This filter effectively eliminates high-frequency content and DC offset from the signal, with a cutoff frequency ranging between 0.5 and 50 Hz. Following the band-pass filter, Independent Component Analysis (ICA) is utilized to remove other unwanted components. ICA is a powerful method that utilizes an iterative algorithm to decompose a group of signals into an orthogonal set of components.

Once the different components have been extracted, the ICA Label Classification tool (also part of EEG-LAB) is utilized to classify these components and determine their sources within the overall EEG signals. This classifier tool categorizes the extracted components into seven different groups, including "brain" (which is the desired signal), "heart," "eye," "muscle," "channel noise," "line noise," and "others." The classifier also provides the probability associated with each component's classification. For instance, two components could both be classified as brain components but with different probabilities. After this classification, any unwanted components (i.e., components other than brain components) are removed, resulting in purified brain EEG signals.

It is important to note that in this study, only components classified as brain components with a probability exceeding 80 percent are selected for rebuilding the decomposed signals. This strict threshold ensures that the signals exclusively represent brain-related EEG signals with a high probability.

### C. Network Data Preperation

Once the noise has been eliminated from the signals, the next step is to generate network data (for the Deep Learning model) using "Forged Channels" method. This method comprises four distinct steps, which we will elaborate on separately to facilitate understanding. The steps involved in this process are "Epoching", "Splitting and Averaging", "SPWVD Analysis", "Resizing and Stacking". In the following, we provide a detailed explanation of each of these steps.

*1) Epoching*

Epoching includes dividing the signal into intervals of N seconds. For instance, if you have a 200-second recording and choose a 2-second epoching, you will obtain 100 epochs. In this study, the duration of epoching is chosen to be 2 seconds. Now since the following steps of preparing the network data is done on each and every epoch, Fig.1 is provided to gives a visual image of how the subsequent steps are done. This figure illustrates the procedure for 1 epoch, starting from the Pre-Processed Signal which obtained its name from the previous stage where data cleaning was conducted. The Pre-Processed signal has a shape of 32 by 1024, meaning it can be visualized as a matrix with 32 rows and 1024 columns. The 32 represents the number of channels in the dataset. The 1024 comes from the fact that the signal was recorded at a sampling frequency of 512, meaning 512 samples were taken every second. So, for a 2-second duration (each epoch was 2 seconds and we are showing 1 epoch here), we have a total of 1024 samples.

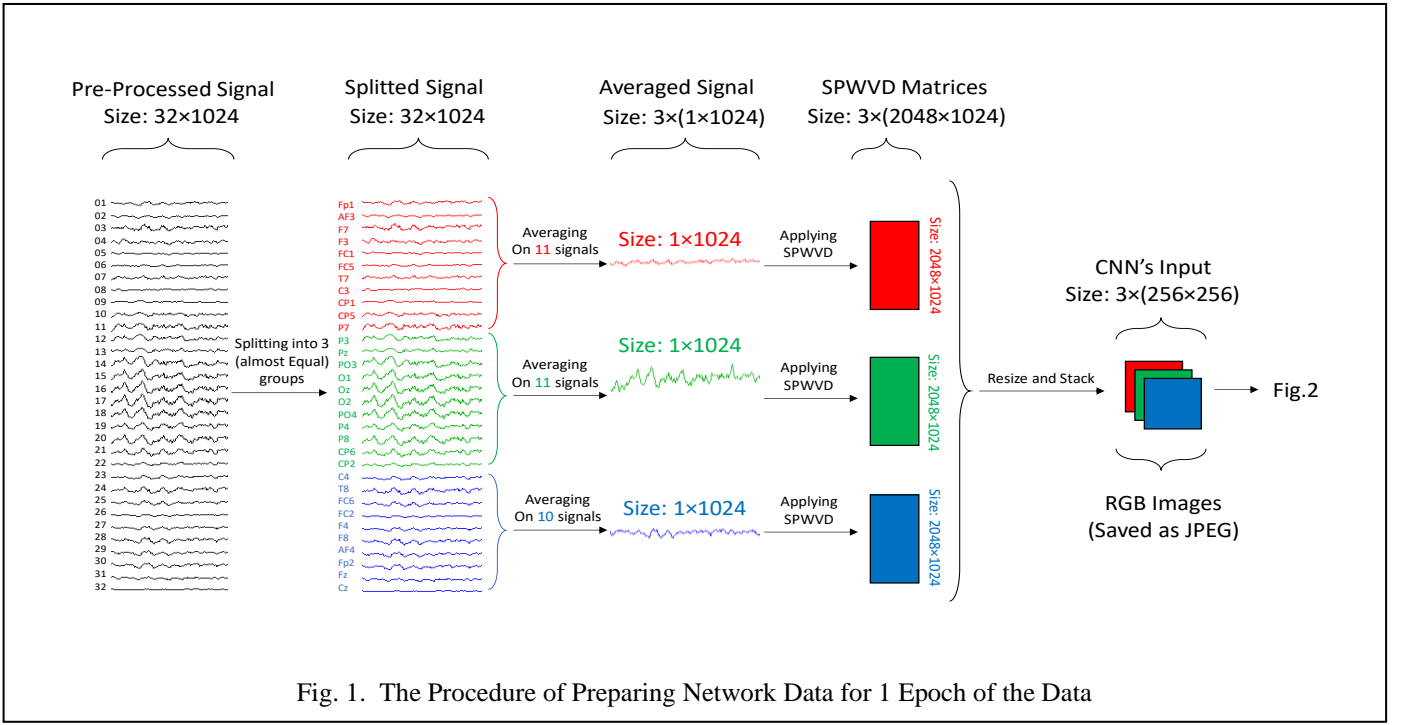

Fig. 1. The Procedure of Preparing Network Data for 1 Epoch of the Data

*2) Splitting and Averaging*

The subsequent step in Network Data Preparation is referred to as splitting. Unlike epoching, splitting is performed on the channels of EEG signals. This process involves dividing the channels of EEG signals into three equal (or almost equal) groups. In this particular dataset, which consists of 32 channels, it is not possible to evenly divide the channels into 3 groups. Consequently, we allocate the first 11 channels to the first group, the next 11 channels to the second group, and the remaining 10 channels to the third group, creating the closest possible division of the 32 channels. The result of splitting process is illustrated in Fig.1 as "Splitted Signal", with the first group represented in red, the second group in green, and the third group in blue. Additionally, once the splitting is completed, the average of each group is calculated. These averaged signals are depicted as the "Averaged Signal" in Fig.1.

Also, the channels in each group are near each other location wise. To illustrate this point, we have written the name of each channel, in the left-hand side of the splitted signals in Fig.1, so that you can conclude the location of all channels in each group.

*3) SPWVD Analysis*

In the subsequent step, a Time-Frequency Representation is applied to these three Averaged signals. In this study, the SPWVD transformation is employed as it is a suitable choice for signals with stochastic characteristics [14]. After applying this transformation, the resulting output will be three matrices, each with a size of 2048×1024 shown in Fig.1 as "SPWVD Matrices".

*4) Resizing and Stacking*

In the final step, each one of the matrices are resized to a dimension of 256×256 and are stacked as an RGB image. By staking we mean putting each one the matrices in one of the channels of an RGB image and saving it as a JPEG file (we used MATLAB to this mean). (An RGB image is a colorful image with 3 channels for the colors red, green and blue)

Some of the Produced images using this method are shown in the Fig. 2. By looking at these images we can conclude several things, explained as follows. First, Since EEG signals are stochastic, the images generated from these signals for the same subject should vary due to changes in the statistics of EEG signals over time. However, as it can be seen, it is evident that the images from different epochs for a given subject are so similar. This indicates that the "Forged Channels" method has effectively addressed the random nature of EEG signals by capturing the deterministic features, resulting in consistent images across epochs for each subject.

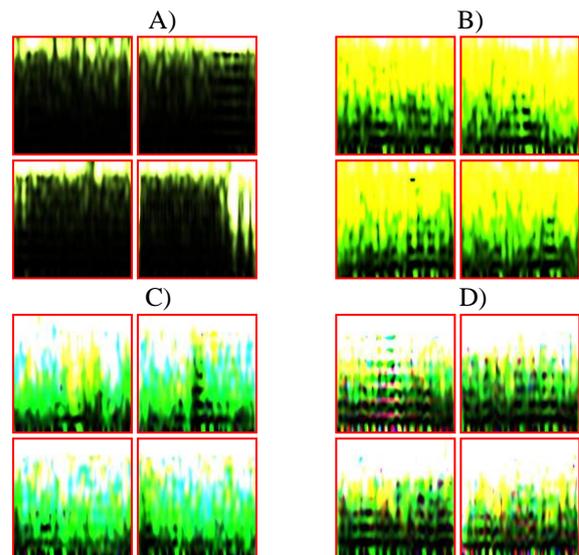

Fig.2. Images Produced using "Forged Channels" method. A) PD subject 1, B) PD subject 4, C) HC subject 16, D) HC subject 7.

It is also worth noting that the images of individuals within the same class exhibit significant differences. This is due to the fact that during the different stage of producing these images, there was no emphasis placed on the specific application at hand, which is the detection of Parkinson's Disease in this case. Consequently, the resulting images serve as a comprehensive representation, or a signature, of an individual's brain activity, and can be utilized in various applications related to EEG signals and brain function.

One last important aspect to note about these images is that they are generated from a 3D matrix, as opposed to a 2D matrix. The issue with using a 2D matrix for plotting colorful images is that they require 3 dimensions, which are lacking in a 2D matrix. Consequently, to compensate for this limitation, we often have to scale the values of the matrix within a random range, which mean adding a random feature to the data which can impact the overall performance of the work. However, with a 3D matrix as input, this problem is avoided entirely, as there is no need for scaling during the image plotting process.

*D. Neural Network.*

Since our dataset is in the form of images, the network used in this study is a Convolutions Neural Network or CNN. CNN's are highly capable of extracting features from the images, and since our images are colorful images, it adds to the fact that these data are now more separable. the summary of our network is shown in the Table I.

As it can be seen, this model contains almost 1.1 million trainable parameters and it was analyzed layer by layer to maximize its efficiency. The specifics for training the model is as follows.

- Adam Optimizer
- Learning rate of 0.0001
- Epoch Size of 30
- Batch Size of 150
- Regularization with Coefficient 0.01 (of Type L2)
- Cross Entropy Cost Function

TABLE I. SUMMARY OF THE USED CNN MODELS

| Layer # - Type | Output Shape | Param # |
|---|---|---|
| 1-Conv2d | [-1, 8, 254, 254] | 224 |
| 2-ReLU | [-1, 8, 254, 254] | 0 |
| 3-MaxPool2d | [-1, 8, 127, 127] | 0 |
| 4-Conv2d | [-1, 16, 125, 25] | 1,168 |
| 5-ReLU | [-1, 16, 125, 25] | 0 |
| 6-Conv2d | [-1, 32, 62, 62] | 4,640 |
| 7-ReLU | [-1, 32, 62, 62] | 0 |
| 8-MaxPool2d | [-1, 32, 31, 31] | 0 |
| 9-Conv2d | [-1, 96, 15, 15] | 27,744 |
| 10-ReLU | [-1, 96, 15, 15] | 0 |
| 11-Flatten | [-1, 21600] | 0 |
| 12-FC | [-1, 50] | 1,080,050 |
| 13-ReLU | [-1, 50] | 0 |
| 14-FC | [-1, 32] | 1,632 |
| 15-ReLU | [-1, 32] | 0 |
| 16-FC | [-1, 2] | 66 |
| 17-Softmax | [-1, 2] | 0 |
| Total Trainable Parameter | | 1,115,524 |

These configurations were chosen based on our available resources (e.g. GPU, RAM) (We have used the free version of Google Colab), and they can be more extreme in case of a better resources.

## III. RESULTS

In this section we want to discuss the results of our work and its performance on the task of classifying HC from PD subject. As it was mentioned before, the type of validation that have been chose to evaluate the performance of the task is Leave-One-subject-Out Cross-Validation.

### A. Leave-One-Subject-Out Cross-Validation

Leave-One-Subject-Out Cross-Validation or LOSOCV is a highly effective method for evaluating machine learning models, particularly in the medical field. Unlike traditional train-test splits where data is randomly shuffled, LOSOCV considers the data of one subject as the test set and the data of all other subjects in the dataset as the training set.
This is what exactly happens in real world situations, in which a model is trained on a dataset (training set) and then being used in practice for new data (test set). Hence this approach closely emulates real-world scenarios, making it a robust form of validation. The results of LOSOCV are presented in the Table II in which the training acc means the accuracy of the model on all subjects except left out subject (i.e. the subject that was mentioned in the second column) and the test acc means the accuracy of the model on the left-out subject.

This table also includes a Pred. column which is decided based on the accuracy of the model on that left out subject. The predictions are denoted by black (true) or red (false) color in the table. It is based on the value of the accuracy and shows how the model can perform in real world. If the accuracy is above 50 percent, indicating that more than half of the subject's data is labeled correctly, it is marked as true. Conversely, if the accuracy is below 50 percent, it is marked as false. Remarkably, approximately 90 percent of the subjects are predicted correctly with a very high certainty. Now given that LOSOCV simulates real-world conditions, this suggests that the model's performance in real-world scenarios is also around 90 percent.

## IV. DISCUSSION

In this section, we will examine the noteworthy observations in table II. One particular observation is the incorrect labeling of three subjects, specifically HC subject 4, PD subject 7, and PD subject 10. These subjects were assigned an accuracy of nearly 0 (or 0) during testing, indicating that the model classified them as belonging to the other class. While the exact cause for this phenomenon may be multifactorial, it is highly likely that these subjects possess a different stochastic distribution in comparison to the other subjects in their own class.

One potential solution to this problem is to increase the size of the dataset. By doing so, the likelihood of encountering a subject with a similar stochastic distribution as the wrongly classified subject also increases, allowing the model to be exposed to a wider range of distributions. With this expanded observation, the model can more accurately identify subjects that exhibit different distributions and perform better in situations where these subjects are present during testing.

TABLE II. RESULT OF THE LOSOCV FOR EACH AND EVERY SUBJECTS

| Type and Number | | Train Loss | Train Acc (%) | Test Loss | Test Acc (%) | Pred. |
|---|---|---|---|---|---|---|
| HC | S01 | 0.412 | 90.58 | 0.371 | 95.83 | HC |
| | S02 | 0.383 | 92.9 | 0.327 | 98.97 | HC |
| | S03 | 0.393 | 92.95 | 0.313 | 100 | HC |
| | S04 | 0.425 | 91.3 | 1.201 | 6.25 | PD |
| | S05 | 0.409 | 90.72 | 0.428 | 90.53 | HC |
| | S06 | 0.415 | 90.43 | 0.313 | 100 | HC |
| | S07 | 0.411 | 90.66 | 0.313 | 100 | HC |
| | S08 | 0.420 | 89.3 | 0.314 | 100 | HC |
| | S09 | 0.406 | 91.06 | 0.316 | 100 | HC |
| | S10 | 0.415 | 89.8 | 0.342 | 96.88 | HC |
| | S11 | 0.411 | 90.6 | 0.328 | 100 | HC |
| | S12 | 0.411 | 90.3 | 0.313 | 100 | HC |
| | S13 | 0.409 | 91.2 | 0.398 | 90.43 | HC |
| | S14 | 0.404 | 90.86 | 0.461 | 87.1 | HC |
| | S15 | 0.408 | 90.92 | 0.360 | 96.94 | HC |
| | S16 | 0.415 | 89.05 | 0.313 | 100 | HC |
| PD | S01 | 0.411 | 90.67 | 0.323 | 100 | PD |
| | S02 | 0.410 | 90.95 | 0.313 | 100 | PD |
| | S03 | 0.416 | 90.21 | 0.336 | 98.95 | PD |
| | S04 | 0.418 | 89.73 | 0.354 | 97.89 | PD |
| | S05 | 0.414 | 90.49 | 0.313 | 100 | PD |
| | S06 | 0.411 | 91.04 | 0.328 | 100 | PD |
| | S07 | 0.392 | 92.57 | 1.313 | 0 | HC |
| | S08 | 0.406 | 91.25 | 0.497 | 83.56 | PD |
| | S09 | 0.414 | 90.02 | 0.407 | 89.36 | PD |
| | S10 | 0.388 | 93.25 | 1.313 | 0 | HC |
| | S11 | 0.411 | 90.97 | 0.411 | 90.22 | PD |
| | S12 | 0.413 | 90.93 | 0.398 | 91.49 | PD |
| | S13 | 0.416 | 90.08 | 0.441 | 90.2 | PD |
| | S14 | 0.409 | 90.65 | 0.442 | 86.32 | PD |
| | S15 | 0.414 | 90.32 | 0.314 | 100 | PD |
| Average | | 0.409 | 90.83 | 0.449 | 86.80 | 90.32 (%) |

TABLE III. COMPARING THE DIFFERENT WORK FROM DIFFERENT ASPECT

| Work | TFR | Accuracy | Validation Type | Computational Cost | Applicability to real world |
|---|---|---|---|---|---|
| [9] | No | 96.95 % | Test Split | Low | Low |
| [10] | No | 88.25 % | Kfold CV (k=10) | Low | Low |
| [11] | No | 82.89 % | Kfold CV(k=5) | Low | Low |
| [7] | Yes | 99.46 % | Kfold CV (k=10) | High | Medium |
| [8] | Yes | 100 % | Kfold CV (k=10) | Medium | Medium |
| [15] | Yes | 70 % | LOSOCV | High | Medium |
| [16] | Yes | 81 % | LOP [17] | High | High |
| This work | Yes | 90.32 % | LOSOCV | Medium | High |

Now with analyzing and elucidating the findings in Table II, we proceed to present Table III, which serves as a comparative analysis between our work and the previous studies mentioned at the beginning of this paper along with some other work that used the same validation setting that we used in this work.

In this table the "TFR" column indicates whether a Time frequency Representation or TFR transformation was employed, while the "Computational Cost" column reflects how many times TFR based method was needed to conduct the data preparation phase based on the mentioned methodology in each paper. Lastly, the "Applicability to the real world" column assesses the suitability of the proposed model based on its accuracy and the strictness of the validation approach chosen for evaluation. The stricter the evaluation method, the more it is applicable to the real world.

As can be seen, when using LOSOCV, the best accuracy is achieved by this work. It is obvious not all the works have used the same evaluation as ours, and therefore no 100 percent conclusion can be made. However, since our type of evaluation is a very strict kind of evaluation, it is safe to say our method is pretty effective in capturing EEG signals features, hence helping in different classification tasks including Parkinson's disease classification.

## V. CONCLUSION

The primary aim of this study was to devise an effective method for obtaining a comprehensive representation of EEG signals to detect PD and HC subject with very high accuracy. The introduction of the innovative "Forged Channels" technique addressed this challenge. Remarkably, this method achieved an impressive classification accuracy of 90.32 percent using LOSOCV evaluation. This result shows that our proposed method is effective in representing EEG signals and is capable of being applied to the real-world scenarios.